\documentstyle{article}
\title{{\bf{ Nonlinear Schr\"{o}dinger Equations and  N=1 Superconformal Algebra}}}
\author{\bf{H.\ T.\"Ozer}\thanks{\bf{e-mail\ :\ ozert @ itu.edu.tr}} and
\bf{S. Saliho\u{g}lu}\thanks{\bf{e-mail\ :\ salihogl  @ itu.edu.tr}} \\\\
 Physics Department,\ Faculty of Science and Letters,\\
Istanbul Technical University,\\
34469,\ Maslak,\ Istanbul,\\
Turkey
}
\begin{document}
\maketitle
\begin{abstract}
\par By using AKNS scheme and  soliton connection taking values in N=1
superconformal algebra we obtain new coupled super Nonlinear
Schr\"odinger equations.
\end{abstract}
\newpage
\setcounter{equation}{0}

\vskip 5mm

\section{\bf{Introduction}}

\par ~~~~~ Using  Ablowitz, Kaub,  Newell , Segur (AKNS) scheme[1] one can
obtain coupled  Nonlinear Schr\"odinger (NLS) equations.Extensions
of coupled NLS equations have been obtained using a simple Lie
algebra[2], a Kac-Moody algebra [3],a Lie superalgebra[3,4],a
Virasoro algebra [5]in the literature.
\par In physics supersymmetry [6] unites bosons and fermions into a
single multiplet. Using supersymmetry one can cancel many normally
divergent Feynman graphs and one can solve the hierarchy problem in
grand unified theories. Supersymmetry also helps to understand the
cosmological constant problem in gravity and it reduces the
divergences of quantum gravity.
\par Conformal invariance in two dimensions is a powerful
symmetry.Two - dimensional quantum field theories that possess
conformal symmetry can be solved exactly by exploiting the conformal
symmetry. Conformal symmetry have found remarkable applications in
string theory and in the study of critical phenomena in statistical
mechanics . N=1 supersymmetric extension of conformal symmetry
,called superconformal symmetry promotes  string to superstring.The
extension of string theory to supersymmetric string theories
identified N=1 superconformal algebra (Neveu-Schwarz type[7] and
Ramond type [8]) as the symmetry algebras of closed superstrings
[9]. The N=1 superconformal algebra with Neveu-Schwarz  and Ramond
types are two possible super - extensions of the Virasoro algebra
[10] for the case of one fermionic current.
\par In  this paper we will obtain super - extensions of coupled NLS
equations using N=1 superconformal algebra with Neveu-Schwarz  and
Ramond types . In sec.2 we will discuss the osp(1,2)superalgebra
valued soliton connection and we will obtain coupled super NLS
equations. Sec.3 and sec.4 concern the soliton connection for the
N=1 superconformal algebra with Neveu-Schwarz type and Ramond type,
respectively and we will obtain in these sections two different
types of super- extensions of coupled NLS equations.


\setcounter{equation}{0}
\section{\bf{ AKNS Scheme with osp(1,2) Superalgebra}}
\par  In  AKNS scheme in 1+1 dimension the  connection is defined as
\begin{equation}
\label{8}
\begin{array}{lll}
\Omega=& \Big(& i \lambda H_1 + Q^{+1} E_{+1}+Q^{-1} E_{-1}+ P^{+{1\over2}} F_{+{1\over2}}+P^{-{1\over2}} F_{-{1\over2}} \Big) dx+\\
& \Big(& -A H_1 + B^{+1} E_{+1}+B^{-1} E_{-1}+ C^{+{1\over2}} F_{+{1\over2}}+C^{-{1\over2}} F_{-{1\over2}} \Big) dt%
\end{array}
\end{equation}
\noindent where  $H_1$, $E_{+1}$, $E_{-1}$are bosonic generators and
$F_{+{1\over2}}$ ,$F_{-{1\over2}}$ are fermionic generators of
osp(1,2) superalgebra. These generators have matrix representations
as
\begin{equation}
\begin{array}{ccc}
H_1 = \left(\begin{array}{cccc}
                               {1\over2} & 0 & 0 \\
                               0 &-{1\over2} & 0 \\
                               0 & 0 & 0 \end{array}
                        \right);
E_{+1} = \left(\begin{array}{cccc}
                               0 & 1 & 0\\
                               0 & 0 & 0\\
                               0 & 0 & 0 \end{array}
                        \right);
E_{-1} = \left(\begin{array}{cccc}
                               0 & 0 & 0\\
                               1 & 0 & 0\\
                               0 & 0 & 0 \end{array}
                        \right)\\
F_{+{1\over2}}= \left(\begin{array}{cccc}
                               0 & 0 & {1\over2} \\
                               0 & 0 & 0 \\
                               0 & {1\over2} & 0 \end{array}
                        \right);
F_{-{1\over2}} = \left(\begin{array}{cccc}
                               0 & 0 & 0 \\
                               0 & 0 & -{1\over2} \\
                               {1\over2} & 0 & 0 \end{array}
                        \right)
\end{array}
\end{equation}
\noindent Also, these generators satisfy the following commutation
and anticommutation  relations
\begin{equation}
\label{17}
\begin{array}{ccc}
 \left[ H_1,E_{\pm 1 }\right] & = &\pm  E_{\pm 1}  \\  \left[E_{+1},E_{-1}\right] & = & 2H_1 \\
 \left[ H_1,F_{\pm {1\over2} }\right] & = &\pm {1\over2} F_{\pm {1\over2}}  \\  \left\{F_{+{1\over2}},F_{-{1\over2}}\right\} & = & {1\over2}H_1\\
  \left[E_{\pm 1 } ,F_{\mp {1\over2}}\right] & = &-  F_{\pm {1\over2} }  \\   \left\{F_{\pm {1\over2}} ,F_{\pm {1\over2}}\right\} & = & \pm {1\over2} E_{\pm 1}\\
\end{array}
\end{equation}
\noindent In Eq.(1) $\lambda$ is the spectral parameter, $Q^{\pm 1}$
and $P^{\pm {1\over2}}$  are fields depending on space and time, namely x
and t, and functions A,$B^{\pm 1}$ and $C^{\pm {1\over2}}$ are x,t and
$\lambda$ dependent.

The integrability condition is given by
\begin{equation}
\label{8}
\begin{array}{lll}
d\ \Omega\  + \ \Omega\ \wedge\ \Omega  =  0
\end{array}
\end{equation}
\par By using Eqs.(1) and (4) one can obtain following equations:
\begin{equation}
\label{18}
\begin{array}{lll}
{Q^{+ 1}}_t=\ \ {B^{+ 1}}_x+  i \lambda  B^{+1} +  Q^{+1} A
+{1\over2}P^{+{1\over2}}C^{+{1\over2}}
\end{array}
\end{equation}
\noindent
\begin{equation}
\label{19}
\begin{array}{lll}
{Q^{- 1}}_t=\ \ {B^{- 1}}_x-  i \lambda  B^{-1} +  Q^{-1} A
-{1\over2}P^{-{1\over2}}C^{-{1\over2}}
\end{array}
\end{equation}
\noindent
\begin{equation}
\label{19}
\begin{array}{lll}
{P^{+{1\over2}}}_t=\ \ {C^{+{1\over2}}}_x+{i\over2} \lambda
C^{+{1\over2}} +{1\over2} P^{+{1\over2}}A
+B^{+1}P^{-{1\over2}}-C^{-{1\over2}}Q^{+1}
\end{array}
\end{equation}
\noindent
\begin{equation}
\label{19}
\begin{array}{lll}
{P^{-{1\over2}}}_t=\ \ {C^{-{1\over2}}}_x-{i\over2} \lambda
C^{-{1\over2}} -{1\over2} P^{-{1\over2}}A
+B^{-1}P^{+{1\over2}}-C^{+{1\over2}}Q^{-1}
\end{array}
\end{equation}
\noindent
\begin{equation}
\label{19}
\begin{array}{lll}
0=A_x+ 2 B^{+ 1}Q^{- 1}-2
B^{-1}Q^{+1}-{1\over2}P^{+{1\over2}}C^{-{1\over2}}-{1\over2}P^{-{1\over2}}C^{+{1\over2}}
\end{array}
\end{equation}
\noindent
\par In AKNS scheme we expand A,$B^{\pm 1}$ and $C^{\pm
{1\over2}}$ in terms of  positive powers of $\lambda$ as
\begin{equation}
A=\sum_{\scriptstyle n=0}^2 \lambda^n a_n;\ \ B^{\pm
1}=\sum_{\scriptstyle n=0}^2 \lambda^n b^{\pm 1}_n;\ \ C^{\pm
{1\over2}}=\sum_{\scriptstyle n=0}^2 \lambda^n c^{\pm{1\over2}}_n
\end{equation}
\noindent Inserting Eq.(10) into Eqs.(5-9)gives 15 relations  in
terms of $a_n$,$b^{\pm 1}_n$ and $c^{\pm {1\over2}}_n$ . By solving
these relations we get
$$a_0=-2 i Q^{+1}Q^{-1}-i P^{+{1\over2}}P^{-{1\over2}};\ \
a_1=0;\ \ a_2=-2 i;\ \
$$
$$
b^{\pm 1}_0=\pm i {Q^{\pm1}}_x ;\
 b^{\pm1}_1= Q^{\pm1};\ b^{\pm1}_2=0
 \eqno(11)
$$
$$
c^{\pm{1\over2}}_0=\pm 2 i {P^{\pm{1\over2}}}_x ;\
 c^{\pm{1\over2}}_1= P^{\pm{1\over2}};\ c^{\pm{1\over2}}_2=0
$$
\noindent By using the relations given by Eq.(11) from Eqs.(5-8) we
obtain the coupled super NLS equations as \setcounter{equation}{11}
\begin{equation}
\label{18}
\begin{array}{lll}
{-i Q^{+1}}_t= {Q^{+1 }}_{xx}-2 (Q^{+1 })^2 Q^{-1}-P^{+{1\over2}}P^{-{1\over2}}Q^{+1}+P^{+{1\over2}}P^{+{1\over2}}_x\\
{~~~i Q^{-1}}_t={Q^{-1 }}_{xx}-2 (Q^{-1})^2Q^{+1}-P^{+{1\over2}}P^{-{1\over2}}Q^{-1}-P^{-{1\over2}}P^{-{1\over2}}_x\\
{-i P^{+{1\over2}}}_t= {2 P^{+{1\over2} }}_{xx}+P^{- {1\over2}}{Q^{+1}}_x+2 Q^{+1}P^{-{1\over2}}_x-Q^{+1}Q^{-1}P^{+{1\over2}}\\
{~~~i P^{-{1\over2}}}_t= {2 P^{-{1\over2} }}_{xx}+P^{+{1\over2} }{Q^{-1}}_x+2 Q^{-1}P^{+{1\over2}}_x-Q^{+1}Q^{-1}P^{-{1\over2}}\\
\end{array}
\end{equation}
\section{\bf{AKNS Scheme with N=1 Superconformal Algebra (Neveu-Schwarz Type)}}
\par We generalize the connection given by Eq.(1) as
\begin{equation}
\label{81}
\begin{array}{lll}
\Omega=&\Big(& i \lambda L_0 + Q^{+m} L_{+m}+Q^{-m} L_{-m}+
P^{+{m\over2}} G_{+{m\over2}}+P^{-{m\over2}} G_{-{m\over2}} \Big) dx+\\
&\Big(& -A L_0 + B^{+m} L_{+m}+B^{-m} L_{-m}+
C^{+{m\over2}} G_{+{m\over2}}+C^{-{m\over2}} G_{-{m\over2}} \Big) dt%
\end{array}
\end{equation}
\noindent where $L_0$ , $L_{\pm m}$ are bosonic generators and
$G_{\pm {m\over2}}$ are fermionic generators of centerless N=1
superconformal algebra of Neveu-Schwarz type , namely they satisfy
the following commutation and anticommutation  relations [6]
\begin{equation}
\label{34}
\begin{array}{lll}
\left[ L_r,L_s\right]&  = & (r-s)\ L_{r+s}\\
\left\{ G_r,G_s\right\}&  = & 2\ L_{r+s}\\
\left[ L_r,G_s\right] & = & ({r\over2}-s)\ G_{r+s}%
\end{array}
\end{equation}
\noindent  Here, $L_{\pm m}$ are generators with positive(negative)
integer indices and $G_{\pm {m\over2}}$ are generators with
positive(negative) half integer indices. In Eq.(13) we assume
summation over the repeated indices. The fields $Q^{\pm m}$ and
 $P^{\pm {m\over2}}$are x,t dependent and also functions A,
 $B^{\pm m}$ and $C^{\pm {m\over2}}$ are x,t and $\lambda$ dependent.
 \vskip 5mm
In N=1 superconformal algebra if we restrict $L_{\pm m}$ to have
only $L_0$ , $L_{\pm 1}$ components and $G_{\pm {m\over2}}$ to have
only $G_{\pm {1\over2}}$ components we obtain osp(1,2) algebra given
by Eq.(3) with the following definitions:
\begin{equation}
\label{342}
\begin{array}{lll}
H_1 =- L_0 ;\ \ E_{+1} =L_{+1} ;\ \ E_{-1} = -L_{-1}\\
 F_{+{1\over2}} ={1\over2}G_{+{1\over2}} ;\ \ F_{-{1\over2}} = -{1\over2}G_{-{1\over2}}%
\end{array}
\end{equation}
\noindent From the integrability condition given by Eq.(4) we obtain
$$
{Q^{+m}}_t= {B^{+m}}_x-i m  \lambda B^{+m}-m A Q^{m}
$$
$$
+\sum_{\scriptstyle r,s=1}^\infty (r-s) B^{+s} Q^{+r} \delta_{r-s,m}
+\sum_{\scriptstyle r,s=1\atop\scriptstyle r>s}^\infty(r+s) B^{-s}
Q^{+r} \delta_{r-s,m}
$$
$$
-\sum_{\scriptstyle r,s=1\atop\scriptstyle r<s}^\infty(r+s) B^{+s}
Q^{-r} \delta_{-r+s,m} +\sum_{\scriptstyle r,s=1\atop\scriptstyle
r>s}^\infty 2 P^{+{r\over2}} C^{-{s\over2}} \delta_{r-s,2
m}\eqno(16)
$$
$$
+\sum_{\scriptstyle r,s=1\atop\scriptstyle r<s}^\infty 2
P^{-{r\over2}} C^{+{s\over2}} \delta_{-r+s,2 m} +\sum_{\scriptstyle
r,s=1\atop\scriptstyle }^\infty 2 P^{+{r\over2}} C^{+{s\over2}}
\delta_{r+s,2 m}
$$
$$
{Q^{-m}}_t= {B^{-m}}_x+i m  \lambda B^{-m}+m A Q^{-m}
$$
$$
-\sum_{\scriptstyle r,s=1}^\infty (r-s) B^{-s} Q^{-r}
\delta_{-r-s,-m} +\sum_{\scriptstyle r,s=1\atop\scriptstyle
r<s}^\infty(r+s) B^{-s} Q^{+r} \delta_{r-s,-m}
$$
$$
-\sum_{\scriptstyle r,s=1\atop\scriptstyle r>s}^\infty(r+s) B^{+s}
Q^{-r} \delta_{-r+s,-m} +\sum_{\scriptstyle r,s=1\atop\scriptstyle
r<s}^\infty 2 P^{+{r\over2}} C^{-{s\over2}} \delta_{r-s,-2
m}\eqno(17)
$$
$$
+\sum_{\scriptstyle r,s=1\atop\scriptstyle r<s}^\infty 2
P^{-{r\over2}} C^{+{s\over2}} \delta_{-r+s,-2 m} +\sum_{\scriptstyle
r,s=1\atop\scriptstyle }^\infty 2 P^{-{r\over2}} C^{-{s\over2}}
\delta_{-r-s,-2 m}
$$
$$
{P^{+{m\over2}}}_t= {C^{+{m\over2}}}_x-{i m\over2}   \lambda
C^{+{m\over2}}-{m\over2}   P^{+{m\over2}} A
$$
$$
+\sum_{\scriptstyle r,s=1\atop\scriptstyle 2 r>s}^\infty
{1\over2}(r+s) Q^{+r} C^{-{s\over2}} \delta_{2r-s,m}
-\sum_{\scriptstyle r,s=1\atop\scriptstyle 2 r<s}^\infty
{1\over2}(r+s) Q^{-r} C^{+{s\over2}} \delta_{-2r+s,m}
$$
$$
+\sum_{\scriptstyle r,s=1}^\infty {1\over2}(r-s) Q^{+r}
C^{+{s\over2}} \delta_{2r+s,m} +\sum_{\scriptstyle
r,s=1\atop\scriptstyle r>2 s}^\infty {1\over2}(r+s)
P^{+{r\over2}}B^{-s} \delta_{r-2 s,m}\eqno(18)
$$
$$
+\sum_{\scriptstyle r,s=1}^\infty {1\over2}(r-s)
P^{+{r\over2}}B^{+s} \delta_{r+2 s,m} -\sum_{\scriptstyle
r,s=1\atop\scriptstyle r<2 s}^\infty {1\over2}(r+s)
P^{-{r\over2}}B^{+s} \delta_{-r+2 s,m}
$$
$$
{P^{-{m\over2}}}_t= {C^{-{m\over2}}}_x+{i m\over2}   \lambda
C^{-{m\over2}}4{m\over2}   P^{-{m\over2}} A
$$
$$
+\sum_{\scriptstyle r,s=1\atop\scriptstyle 2 r<s}^\infty
{1\over2}(r+s) Q^{+r} C^{-{s\over2}} \delta_{2r-s,-m}
-\sum_{\scriptstyle r,s=1\atop\scriptstyle 2 r>s}^\infty
{1\over2}(r+s) Q^{-r} C^{+{s\over2}} \delta_{-2r+s,-m}
$$
$$
-\sum_{\scriptstyle r,s=1}^\infty {1\over2}(r-s) Q^{-r}
C^{-{s\over2}} \delta_{-2r-s,-m} +\sum_{\scriptstyle
r,s=1\atop\scriptstyle r<2 s}^\infty {1\over2}(r+s)
P^{+{r\over2}}B^{-s} \delta_{r-2 s,-m}\eqno(19)
$$
$$
-\sum_{\scriptstyle r,s=1}^\infty {1\over2}(r-s)
P^{-{r\over2}}B^{-s} \delta_{-r-2 s,-m} -\sum_{\scriptstyle
r,s=1\atop\scriptstyle r>2 s}^\infty {1\over2}(r+s)
P^{-{r\over2}}B^{+s} \delta_{-r+2 s,-m}
$$
\noindent and
$$
0=-A_x+2 \sum_{\scriptstyle r=1}^\infty \left(
   r B^{-r} Q^{+r}
 - r B^{+r} Q^{-r}
 + P^{+{r\over2}} C^{-{r\over2}}
 + P^{-{r\over2}} C^{+{r\over2}}\right)\eqno(20)
$$
\vskip 3mm
\par In AKNS scheme we expand A,$B^{\pm m}$ and $C^{\pm
{m\over2}}$ in terms of the positive powers of $\lambda$ as
\setcounter{equation}{20}

\begin{equation}
A=\sum_{\scriptstyle n=0}^2 \lambda^n a_n;\ \ B^{\pm
m}=\sum_{\scriptstyle n=0}^2 \lambda^n b^{\pm m}_n;\ \ C^{\pm
{m\over2}}=\sum_{\scriptstyle n=0}^2 \lambda^n c^{\pm{m\over2}}_n
\end{equation}
\noindent Inserting Eq.(21) into Eqs.(16-20)gives 15 relations  in
terms of $a_n$,$b^{\pm m}_n$ and $c^{\pm {m\over2}}_n$ (n=0,1,2) .
By solving these relations we get

$$a_0=2 i \sum_{\scriptstyle r=1}^\infty Q^{+r}Q^{-r}
+{{4i}\over m}\sum_{\scriptstyle r=1}^\infty
P^{+{r\over2}}P^{-{r\over2}};\ a_1=const.=a_{10};\ a_2=- i;\ \
$$
$$
b^{\pm m}_0=\mp {i\over m} {Q^{\pm m}}_x + a_{10} i Q^{\pm m} ;\
 b^{\pm m}_1= Q^{\pm m};\ b^{\pm m}_2=0
 \eqno(22)
$$
$$
c^{\pm {m\over2}}_0= a_{10} i P^{\pm {m\over2}}\mp {{2i}\over m}
{P^{\pm {m\over2} }}_x  ;\
 c^{\pm {m\over2}}_1= P^{\pm {m\over2}};\ c^{\pm {m\over2}}_2=0
$$

\noindent By using the relations given by Eq.(22) from Eqs.(16-19)
we obtain the coupled super NLS equations as
$$
{Q^{+m}}_t={-{i}\over{m}}{Q^{+m}}_{xx}+i a_{10} {Q^{+m}}_{x}
$$
$$
 -2 i m
Q^{+m}\left(\sum_{\scriptstyle r=1}^\infty Q^{+r}Q^{-r}\right)
 -4 i
m Q^{+m}\left(\sum_{\scriptstyle
r=1}^\infty\left[{{1}\over{r}}\right]
P^{+{r\over2}}P^{-{r\over2}}\right)
$$
$$
-i \sum_{\scriptstyle r=1\atop\scriptstyle
r<m}^\infty\left[{{2r-m}\over{m-r}}\right]Q^{+r}Q^{+(m-r)}_x
-i\sum_{\scriptstyle r=1\atop\scriptstyle
r>m}^\infty\left[{{2r-m}\over{m-r}}\right]Q^{+r}Q^{-(r-m)}_x
$$
$$
+i\sum_{\scriptstyle r=1\atop\scriptstyle
r>m}^\infty\left[{{2r-m}\over{r}}\right]Q^{-(r-m)}Q^{+r}_x
+4i\sum_{\scriptstyle r=1\atop\scriptstyle
r>2m}^\infty\left[{{1}\over{r-2m}}\right]P^{+{r\over2}}
P^{-({r\over2}-m)}_x\eqno(23)
$$
$$
-4i\sum_{\scriptstyle r=1\atop\scriptstyle
r>2m}^\infty\left[{{1}\over{r}}\right]P^{-({r\over2}-m)}P^{+{r\over2}}_x
-4i\sum_{\scriptstyle r=1\atop\scriptstyle
r<2m}^\infty\left[{{1}\over{2m-r}}\right]P^{+{r\over2}}P^{+(m-{r\over2})}_x
$$

\noindent
$$
{Q^{-m}}_t={{i}\over{m}}{Q^{-m}}_{xx}+i a_{10} {Q^{-m}}_{x}
$$
$$
 +2 i m
Q^{-m}\left(\sum_{\scriptstyle r=1}^\infty Q^{+r}Q^{-r}\right)
 +4 i
m Q^{-m}\left(\sum_{\scriptstyle
r=1}^\infty\left[{{1}\over{r}}\right]
P^{+{r\over2}}P^{-{r\over2}}\right)
$$
$$
+i \sum_{\scriptstyle r=1\atop\scriptstyle
r<m}^\infty\left[{{2r-m}\over{r-m}}\right]Q^{-r}Q^{+(r-m)}_x
+i\sum_{\scriptstyle r=1\atop\scriptstyle
r>m}^\infty\left[{{2r-m}\over{r-m}}\right]Q^{-r}Q^{+(r-m)}_x
$$
$$
+i\sum_{\scriptstyle r=1\atop\scriptstyle
r>m}^\infty\left[{{2r-m}\over{r}}\right]Q^{+(r-m)}Q^{-r}_x
+4i\sum_{\scriptstyle r=1\atop\scriptstyle
r>2m}^\infty\left[{{1}\over{r}}\right]P^{+({{r\over2}-m})}P^{-{r\over2}}_x\eqno(24)
$$
$$
-4i\sum_{\scriptstyle r=1\atop\scriptstyle
r>2m}^\infty\left[{{1}\over{r-2m}}\right]P^{-{r\over2}}P^{+({{r\over2}-m})}_x
-4i\sum_{\scriptstyle r=1\atop\scriptstyle
r<2m}^\infty\left[{{1}\over{r-2m}}\right]P^{-{r\over2}}P^{+({{r\over2}-m})}_x
$$
$$
{P^{+{m\over2}}}_t={-{2i}\over{m}}{P^{+{m\over2}}}_{xx}+i a_{10}
{P^{+{m\over2}}}_{x}
$$
$$
 - i m
{P^{+{m\over2}}}\left(\sum_{\scriptstyle r=1}^\infty
Q^{+r}Q^{-r}\right)
 -2 i
m {P^{+{m\over2}}}\left(\sum_{\scriptstyle
r=1}^\infty\left[{{1}\over{r}}\right]
P^{+{r\over2}}P^{-{r\over2}}\right)
$$
$$
+{i\over2}\sum_{\scriptstyle r=1\atop\scriptstyle
r>m}^\infty\left[{{3r-m}\over{r-m}}\right]P^{+{r\over2}}Q^{-({r\over2}-{m\over2})}_x
+{i\over2}\sum_{\scriptstyle r=1\atop\scriptstyle
r>{m\over2}}^\infty\left[{{3r-m}\over{r}}\right]P^{-(r-{m\over2})}
Q^{+{r}}_x
$$
$$
+{i\over2}\sum_{\scriptstyle r=1\atop\scriptstyle
r<m}^\infty\left[{{3r-m}\over{r-m}}\right]P^{+{r\over2}}Q^{-({m\over2}-{r\over2})}_x
+{i}\sum_{\scriptstyle r=1\atop\scriptstyle
r>{m\over2}}^\infty\left[{{3r-m}\over{2r-m}}\right]Q^{+{r\over2}}P^{-(r-{m\over2})}_x\eqno(25)
$$
$$
+{i}\sum_{\scriptstyle r=1\atop\scriptstyle
r>m}^\infty\left[{{3r-m}\over{2r}}\right]Q^{-({r\over2}-{m\over2})}
P^{+{r\over2}}_x +{i}\sum_{\scriptstyle r=1\atop\scriptstyle
r<{m\over2}}^\infty\left[{{3r-m}\over{2r-m}}\right]Q^{+{r}}P^{-({m\over2}-r)}_x
$$
\noindent and
$$
{P^{-{m\over2}}}_t={{2i}\over{m}}{P^{-{m\over2}}}_{xx}+i a_{10}
{P^{-{m\over2}}}_{x}
$$
$$
 + i m
{P^{-{m\over2}}}\left(\sum_{\scriptstyle r=1}^\infty
Q^{+r}Q^{-r}\right)
 +2 i
m {P^{-{m\over2}}}\left(\sum_{\scriptstyle
r=1}^\infty\left[{{1}\over{r}}\right]
P^{+{r\over2}}P^{-{r\over2}}\right)
$$
$$
+{i\over2}\sum_{\scriptstyle r=1\atop\scriptstyle
r>{m\over2}}^\infty\left[{{3r-m}\over{r}}\right]P^{+(r-{m\over2})}
Q^{-{r}}_x +{i\over2}\sum_{\scriptstyle r=1\atop\scriptstyle
r>m}^\infty\left[{{3r-m}\over{r-m}}\right]P^{-{r\over2}}Q^{+({r\over2}-{m\over2})}_x
$$
$$
+{i\over2}\sum_{\scriptstyle r=1\atop\scriptstyle
r<m}^\infty\left[{{3r-m}\over{r-m}}\right]P^{-{r\over2}}Q^{+({r\over2}-{m\over2})}_x
+{i}\sum_{\scriptstyle r=1\atop\scriptstyle
r>m}^\infty\left[{{3r-m}\over{2r}}\right]Q^{+({r\over2}-{m\over2})}
P^{-{r\over2}}_x\eqno(26)
$$
$$
+{i}\sum_{\scriptstyle r=1\atop\scriptstyle
r>{m\over2}}^\infty\left[{{3r-m}\over{2r-m}}\right]Q^{-r}P^{+(r-{m\over2})}_x
+{i}\sum_{\scriptstyle r=1\atop\scriptstyle
r<{m\over2}}^\infty\left[{{3r-m}\over{2r-m}}\right]Q^{-{r}}P^{+(r-{m\over2})}_x
$$

\section{\bf{AKNS Scheme with N=1 Superconformal Algebra (Ramond Type)}}
\setcounter{equation}{26}
\par We take the soliton  connection as
\begin{equation}
\label{81}
\begin{array}{lll}
\Omega=&\Big(& i \lambda L_0 + Q^{+m} L_{+m}+Q^{-m} L_{-m}+
P^{+m} G_{+ m}+P^{-m} G_{-m} \Big) dx+\\
&\Big(& A L_0 + D G_0+ B^{+m} L_{+m}+B^{-m} L_{-m}+
C^{+m} G_{+m}+C^{-m} G_{-m} \Big) dt%
\end{array}
\end{equation}
\noindent  where $L_0$ , $L_{\pm m}$ are bosonic generators and
$G_{\pm {m}}$ are fermionic generators of centerless N=1
superconformal algebra of Ramond type , namely they satisfy the
following commutation and anticommutation  relations [6]
\begin{equation}
\label{34}
\begin{array}{lll}
\left[ L_r,L_s\right]&  = & (r-s)\ L_{r+s}\\
\left\{ G_r,G_s\right\}&  = & 2\ L_{r+s}\\
\left[ L_r,G_s\right] & = & ({r\over2}-s)\ G_{r+s}%
\end{array}
\end{equation}
\noindent  Here,$L_{\pm m}$ and $G_{\pm m}$are generators with
positive(negative) integer indices. In Eq.(27) we assume summation
over the repeated indices. The fields $Q^{\pm m}$ and
 $P^{\pm m}$are x,t dependent and also functions A,D,
 $B^{\pm m}$ and $C^{\pm m}$ are x,t and $\lambda$ dependent.
 \vskip 5mm
\noindent From the integrability condition given by Eq.(4) we obtain
$$
{Q^{+m}}_t= {B^{+m}}_x-i m  \lambda B^{+m}-m A Q^{m}-2 P^{+m} D
$$
$$
+\sum_{\scriptstyle r,s=1}^\infty \left(2 P^{+s} C^{+r}+(s-r) Q^{+s}
B^{+r}\right) \delta_{r+s,m}
$$
$$
+\sum_{\scriptstyle r,s=1\atop\scriptstyle r<s}^\infty \left(2
P^{-r} C^{+s}-(r+s) Q^{-r} B^{+s}\right) \delta_{-r+s,m}\eqno(29)
$$
$$
+\sum_{\scriptstyle r,s=1\atop\scriptstyle r>s}^\infty \left(2
P^{+r} C^{-s}+(r+s) Q^{+r} B^{-s}\right) \delta_{r-s,m}
$$
$$
{Q^{-m}}_t= {B^{-m}}_x+i m  \lambda B^{-m}+m A Q^{-m}-2 P^{-m} D
$$
$$
+\sum_{\scriptstyle r,s=1}^\infty \left(2 P^{+s} C^{+r}+(s-r) Q^{+s}
B^{+r}\right) \delta_{r+s,m}
$$
$$
+\sum_{\scriptstyle r,s=1\atop\scriptstyle r<s}^\infty \left(2
P^{-r} C^{+s}-(r+s) Q^{-r} B^{+s}\right) \delta_{-r+s,m}\eqno(30)
$$
$$
+\sum_{\scriptstyle r,s=1\atop\scriptstyle r>s}^\infty \left(2
P^{+r} C^{-s}+(r+s) Q^{+r} B^{-s}\right) \delta_{r-s,m}
$$
$$
{P^{+m}}_t= {C^{+m}}_x-i m  \lambda C^{+m}-m P^{+m}
A-{m\over2}Q^{+m} D
$$
$$
-{1\over2}\sum_{\scriptstyle r,s=1}^\infty \left(\left({{r-2
s}}\right) P^{+r} B^{+s}+\left({{2r-s}}\right) Q^{+s} C^{+r}\right)
\delta_{r+s,m}
$$
$$
-{1\over2}\sum_{\scriptstyle r,s=1\atop\scriptstyle r<s}^\infty
\left(\left({{2r+ s}}\right) P^{-r} B^{+s}+\left({{r+2 s}}\right)
Q^{-r} C^{+s}\right) \delta_{-r+s,m}\eqno(31)
$$
$$
+{1\over2}\sum_{\scriptstyle r,s=1\atop\scriptstyle r>s}^\infty
\left(\left({{2r+ s}}\right) P^{+r} B^{-s}+\left({{r+2 s}}\right)
Q^{+r} C^{-s}\right) \delta_{r-s,m}
$$

$$
{P^{-m}}_t= {C^{-m}}_x+i m  \lambda C^{-m}+m P^{-m}
A+{m\over2}Q^{-m} D
$$
$$
{1\over2}\sum_{\scriptstyle r,s=1}^\infty \left(\left({{s-2
r}}\right) P^{-r} B^{-s}+\left({{s-2r}}\right) Q^{-r} C^{-s}\right)
\delta_{-r-s,-m}
$$
$$
-{1\over2}\sum_{\scriptstyle r,s=1\atop\scriptstyle r>s}^\infty
\left(\left({{r+ 2s}}\right) P^{-r} B^{+s}+\left({{r+2 s}}\right)
Q^{-r} C^{+s}\right) \delta_{-r+s,-m}\eqno(32)
$$
$$
+{1\over2}\sum_{\scriptstyle r,s=1\atop\scriptstyle r<s}^\infty
\left(\left({{2r+ s}}\right) P^{+r} B^{-s}+\left({{r+2 s}}\right)
Q^{+r} C^{-s}\right) \delta_{r-s,-m}
$$
$$
0=-A_x+2 \sum_{\scriptstyle r=1}^\infty \left(  r B^{-r} Q^{+r}
 - r B^{+r} Q^{-r} + P^{+r} C^{-r} + P^{-r} C^{+r}\right)\eqno(33)
$$
\noindent and
$$
0=-D_x+{3\over2} \sum_{\scriptstyle r=1}^\infty r \left(  P^{+r}
B^{-r}
 - P^{-r} B^{+r} - Q^{-r} C^{+r} + Q^{+r} C^{-r}\right)\eqno(34)
$$
\vskip 3mm
\par In AKNS scheme we expand A,D,$B^{\pm m}$ and $C^{\pm
m}$ in terms of the positive powers of $\lambda$ as
\setcounter{equation}{34}

\begin{equation}
A=\sum_{\scriptstyle n=0}^2 \lambda^n a_n;\ \ D=\sum_{\scriptstyle
n=0}^2 \lambda^n d_n;\ \ B^{\pm m}=\sum_{\scriptstyle n=0}^2
\lambda^n b^{\pm m}_n;\ \ C^{\pm m}=\sum_{\scriptstyle n=0}^2
\lambda^n c^{\pm m}_n
\end{equation}
\noindent Inserting Eq.(35) into Eqs.(29-34)gives 18 relations  in
terms of $a_n$,$d_n$,$b^{\pm m}_n$ and $c^{\pm m}_n$ (n=0,1,2) . By
solving these relations we get

$$a_0=2 i \sum_{\scriptstyle r=1}^\infty Q^{+r}Q^{-r}
+{{2i}\over m}\sum_{\scriptstyle r=1}^\infty P^{+r}P^{-r};\ a_1=0;\
a_2=- i;\ \
$$
$$
b^{\pm m}_0=\mp {i\over m} {Q^{\pm m}}_x ;\
 b^{\pm m}_1= Q^{\pm m};\ b^{\pm m}_2=0
 \eqno(36)
$$
$$
c^{\pm m}_0=\mp {{i}\over m} {P^{\pm m }}_x ;\
 c^{\pm m}_1= P^{\pm m};\ c^{\pm m}_2=0
$$
$$d_0=
 {{3i}\over 2}\sum_{\scriptstyle r=1}^\infty P^{-r}Q^{+r}
+{{3i}\over 2}\sum_{\scriptstyle r=1}^\infty Q^{+r}P^{-r} ;\ d_1=0;\
d_2=0;\ \
$$
\noindent By using the relations given by Eq.(36) from Eqs.(29-32)
we obtain the coupled super NLS equations as
$$
{Q^{+m}}_t={-{i}\over{m}}{Q^{+m}}_{xx} -2 i m
Q^{+m}\left(\sum_{\scriptstyle r=1}^\infty Q^{+r}Q^{-r}\right)
$$
$$
-4 i m Q^{+m}\left(\sum_{\scriptstyle
r=1}^\infty\left[{{1}\over{r}}\right]
P^{+{r\over2}}P^{-{r\over2}}\right)
-3 i\sum_{\scriptstyle
r=1}^\infty\left( Q^{+r}P^{-r}+Q^{-r}P^{+r}\right)
$$
$$
-i \sum_{\scriptstyle r=1\atop\scriptstyle r<m}^\infty\left(
\left[{{2}\over{r}}\right]
 P^{+(m-r)}P^{+r}_x-\left[{{2r-m}\over{r}}\right]Q^{+(m-r)}Q^{+r}_x\right)\eqno(37)
$$
$$
+i \sum_{\scriptstyle r=1\atop\scriptstyle r>m}^\infty\left(
\left[{{2}\over{r-m}}\right]
 P^{+r}P^{-(r-m)}_x+\left[{{2r-m}\over{r-m}}\right]Q^{+r}Q^{-(r-m)}_x\right)
$$
$$
-i \sum_{\scriptstyle r=1\atop\scriptstyle r>m}^\infty\left(
\left[{{2}\over{r}}\right]
 P^{-(r-m)}P^{+r}_x-\left[{{2r-m}\over{r}}\right]Q^{-(r-m)}Q^{+r}_x\right)
$$
$$
{Q^{-m}}_t={{i}\over{m}}{Q^{-m}}_{xx} +2 i m
Q^{-m}\left(\sum_{\scriptstyle r=1}^\infty Q^{+r}Q^{-r}\right)
$$
$$
+4 i m Q^{-m}\left(\sum_{\scriptstyle
r=1}^\infty\left[{{1}\over{r}}\right]
P^{+{r\over2}}P^{-{r\over2}}\right) +3 i\sum_{\scriptstyle
r=1}^\infty\left( Q^{+r}P^{-r}+Q^{-r}P^{+r}\right)
$$
$$
-i \sum_{\scriptstyle r=1\atop\scriptstyle r<m}^\infty\left(
\left[{{2}\over{r-m}}\right]
 P^{-r}P^{+(r-m)}_x+\left[{{2r-m}\over{r-m}}\right]Q^{-r}Q^{+(r-m)}_x\right)\eqno(38)
$$
$$
+i \sum_{\scriptstyle r=1\atop\scriptstyle r>m}^\infty\left(
\left[{{2}\over{r}}\right]
 P^{+(r-m)}P^{-r}_x+\left[{{2r-m}\over{r}}\right]Q^{+(r-m)}Q^{-r}_x\right)
$$
$$
-i \sum_{\scriptstyle r=1\atop\scriptstyle r>m}^\infty\left(
\left[{{2}\over{r-m}}\right]
 P^{-r}P^{+(r-m)}_x-\left[{{2r-m}\over{r-m}}\right]Q^{-r}Q^{+(r-m)}_x\right)
$$
$$
{P^{+m}}_t={-{i}\over{m}}{P^{+m}}_{xx} -2 i m
P^{+m}\left(\sum_{\scriptstyle r=1}^\infty Q^{+r}Q^{-r}\right)
$$
$$
-2 i m P^{+m}\left(\sum_{\scriptstyle
r=1}^\infty\left[{{1}\over{r}}\right]
P^{+{r\over2}}P^{-{r\over2}}\right) -{3\over4} i m
Q^{+m}\sum_{\scriptstyle r=1}^\infty\left(
Q^{+r}P^{-r}+Q^{-r}P^{+r}\right)
$$
$$
+{i\over 2} \sum_{\scriptstyle r=1\atop\scriptstyle
r<m}^\infty\left( \left[{{2 r-m}\over{r}}\right]
 P^{+(m-r)}Q^{+r}_x+\left[{{r-m}\over{r}}\right]Q^{+(m-r)}P^{+r}_x\right)\eqno(39)
$$
$$
+{i\over 2} \sum_{\scriptstyle r=1\atop\scriptstyle
r>m}^\infty\left( \left[{{3 r-m}\over{r-m}}\right]
 P^{+r}Q^{-(r-m)}_x+\left[{{3 r-2 m}\over{r-m}}\right]Q^{+r}P^{-(r-m)}_x\right)
$$
$$
+{i\over 2} \sum_{\scriptstyle r=1\atop\scriptstyle
r>m}^\infty\left( \left[{{3 r-m}\over{r}}\right]
 P^{-(r-m)}Q^{+r}_x+\left[{{3 r-m}\over{r}}\right]Q^{-(r-m)}P^{+r}_x\right)
$$

$$
{P^{-m}}_t={{i}\over{m}}{P^{-m}}_{xx} +2 i m
P^{-m}\left(\sum_{\scriptstyle r=1}^\infty Q^{+r}Q^{-r}\right)
$$
$$
+2 i m P^{-m}\left(\sum_{\scriptstyle
r=1}^\infty\left[{{1}\over{r}}\right]
P^{+{r\over2}}P^{-{r\over2}}\right) +{3\over4} i m
Q^{-m}\sum_{\scriptstyle r=1}^\infty\left(
Q^{+r}P^{-r}+Q^{-r}P^{+r}\right)
$$
$$
+{i\over 2} \sum_{\scriptstyle r=1\atop\scriptstyle
r<m}^\infty\left( \left[{{3 r-m}\over{r-m}}\right]
 P^{-r}Q^{+(r-m)}_x+\left[{{3r-2m}\over{r-m}}\right]Q^{-r}P^{+(r-m)}_x\right)\eqno(40)
$$
$$
+{i\over 2} \sum_{\scriptstyle r=1\atop\scriptstyle
r>m}^\infty\left( \left[{{3 r-m}\over{r}}\right]
 P^{+(r-m)}Q^{-r}_x+\left[{{3 r- m}\over{r}}\right]Q^{+(r-m)}P^{-r}_x\right)
$$
$$
+{i\over 2} \sum_{\scriptstyle r=1\atop\scriptstyle
r>m}^\infty\left( \left[{{3 r-m}\over{r-m}}\right]
 P^{-r}Q^{+(r-m)}_x+\left[{{3 r-m}\over{r-m}}\right]Q^{-r}P^{+(r-m)}_x\right)
$$
\section{\bf{Conclusions}}
\par ~~~~~ Using AKNS scheme and N=1 superconformal algebra of
Neveu-Schwarz and Ramond types we obtain two different new super-
extensions of  coupled Nonlinear Schr\"odinger equations.


\end{document}